\documentclass[Physsubmission, Phys]{SciPost}
\hypersetup{
    colorlinks,
    linkcolor={red!50!black},
    citecolor={blue!50!black},
    urlcolor={blue!80!black}
}

\usepackage[bitstream-charter]{mathdesign}
\DeclareSymbolFont{usualmathcal}{OMS}{cmsy}{m}{n}
\DeclareSymbolFontAlphabet{\mathcal}{usualmathcal}

\usepackage{xspace}
\newcommand{\HEJ}{\texttt{HEJ}\xspace}

\newcommand{\HEJNLO}{\texttt{HEJ}\text{@NLO}\xspace}

\usepackage{caption}
\usepackage{subcaption}

\begin{document}

% TODO: write your article's title here.
% The article title is centered, Large boldface, and should fit in two lines
\begin{center}{\Large \textbf{
Logarithmic corrections for jet production at the LHC\\
}}\end{center}

% TODO: write the author list here. Use initials + surname format.
% Separate subsequent authors by a comma, omit comma at the end of the list.
% Mark the corresponding author with a superscript *.
\begin{center}
%J. R. Andersen\textsuperscript{1},
E. P. Byrne\textsuperscript{1$\star$}
%A. Maier\textsuperscript{3} and
%J. M. Smillie\textsuperscript{2}
\end{center}

% TODO: write all affiliations here.
% Format: institute, city, country
\begin{center}
%{\bf 1} Institute for Particle Physics Phenomenology, University of Durham, Durham, DH1 3LE, UK
%\\
{\bf 1} Higgs Centre for Theoretical Physics, University of Edinburgh,\\
  Peter Guthrie Tait Road, Edinburgh, EH9 3FD, UK
\\
%{\bf 3} Deutsches Elektronen-Synchrotron, DESY, Platanenallee 6,
%  15738 Zeuthen, Germany
%\\
% TODO: provide email address of corresponding author
* emmet.byrne@ed.ac.uk
\end{center}

\begin{center}
\today
\end{center}

% For convenience during refereeing (optional),
% you can turn on line numbers by uncommenting the next line:
%\linenumbers
% You should run LaTeX twice in order for the line numbers to appear.

\definecolor{palegray}{gray}{0.95}
\begin{center}
\colorbox{palegray}{
  \begin{tabular}{rr}
  \begin{minipage}{0.1\textwidth}
    \includegraphics[width=30mm]{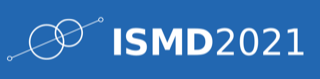}
  \end{minipage}
  &
  \begin{minipage}{0.75\textwidth}
    \begin{center}
    {\it 50th International Symposium on Multiparticle Dynamics}\\ {\it (ISMD2021)}\\
    {\it 12-16 July 2021} \\
    \doi{10.21468/SciPostPhysProc.?}\\
    \end{center}
  \end{minipage}
\end{tabular}
}
\end{center}

\section*{Abstract}
{\bf
% TODO: write your abstract here.
Several important processes and analyses at the LHC are sensitive to higher-order perturbative corrections beyond what can currently be calculated at fixed order. One important class of logarithmic corrections are those which appear when the centre-of-mass energy of a QCD collision is much larger than the transverse momenta of the observed jets.
We describe the High Energy Jets (\HEJ) framework, which includes the dominant high-energy logarithms to provide all-order predictions for several LHC processes including Higgs, $W$, or $Z$ boson production in association with jets. We will summarise some recent developments, in particular the first matching of \HEJ to a NLO calculation.
}

% TODO: include a table of contents (optional)
% Guideline: if your paper is longer that 6 pages, include a TOC
% To remove the TOC, simply cut the following block
%\vspace{10pt}
%\noindent\rule{\textwidth}{1pt}
%\tableofcontents\thispagestyle{fancy}
%\noindent\rule{\textwidth}{1pt}
%\vspace{10pt}

\section{Introduction}
\label{sec:intro}
Standard Model predictions for the LHC are typically produced using general purpose Monte Carlo event generators which describe many aspects of a proton-proton collision, from PDFs to hadronisation. Here we restrict our attention to the description of the hard matrix element which describes the scattering between one parton from each proton. The hard matrix element is typically computed via fixed-order perturbation theory. 
For QCD amplitudes, the coupling $\alpha_s$ decreases with increasing centre of mass energy.
However, there are regions of phase space where an expansion in $\alpha_s$ is unstable. For example, consider the cross section for 2$\to$2 scattering of partons in the region where the Mandelstam variable $s$ is much larger than $|t|$. At order $\alpha_s^{n+2}$ there are terms which are of the form $\log^n(s/|t|)$. At energies accessible by the LHC the quantity $\left( \alpha_s \log(s/|t|)\right)^n$ can be of order one. Logarithms of this type are called high-energy logarithms and must be summed to all orders to ensure stability of our perturbative predictions. This motivates an alternative classification of accuracy, where N$^m$LL refers to all terms in the perturbative series of the form $\alpha_s^{n+2} \log(s/|t|)^{n-m}$. 
\par
The LL contribution to QCD amplitudes, to all orders in $\alpha_s$, takes a simple factorised form \cite{Fadin:1975cb,Lipatov:1976zz,Kuraev:1977fs}. Rather than describing the hard matrix element of a collision at the LHC via fixed-order perturbation theory we can instead take as our starting point matrix elements which are  LL accurate. High Energy Jets (\HEJ) is a framework that builds on this LL accuracy. However, \HEJ utilises Monte Carlo integration for phase space integration, which means minimal approximations to the amplitude need to be made, and no formal approximations to phase space. This in turn allows HEJ to make all-order predictions for not only asymptotically large energies, but also at the scales of energy accessible by the LHC and for arbitrary cuts and analyses.
\par
In Section \ref{sec:factorisation} of we review the simple factorised form that amplitudes take in the high-energy limit. In Section \ref{sec:HEJ} we show how the \HEJ framework builds upon this factorised picture to obtain an improved LL accurate description, and we discuss NLL improvements to this framework. In Section \ref{sec:matching} we discuss how this framework can be matched to existing fixed-order calculations. We review a recent application of this framework to the production of a $W$ boson in association with at least two jets before stating our conclusions.
\section{Factorisation of amplitudes in the high-energy limit}
\label{sec:factorisation}
To LL accuracy, the amplitudes for $2 \to n$ processes in QCD (with $n\ge 2$) take a simple factorised form \cite{Fadin:1975cb,Lipatov:1976zz,Kuraev:1977fs}. These amplitudes become exact in the so-called Multi-Regge-Kinematic (MRK) limit, where for momenta $p_a,p_b\to p_1,\cdots, p_n$ the outgoing momenta are strictly ordered in rapidity $y_i$, while the transverse momenta $\mathbf{p}_i$ are of the same finite magnitude:
\begin{align}
\label{eq:MRK}
\begin{split}
    &y_{i} \gg y_{i+1} \quad \forall i \in \{1, \cdots, n-1\}, \quad |\mathbf{p}_i|\approx|\mathbf{p}_j| \quad \forall i,j \in \{1, \cdots, n\}.
\end{split}
\end{align}
As an illustration of the simplicity of these amplitudes, we give the squared amplitude for the scattering of distinguishable quarks $q$ and $Q$ to $n$ partons:
\begin{align}
\label{eq:LLMRK}
\begin{split}
    \left|\mathcal{M}^{\mathrm{LL \ MRK}}_{qQ\to n}\right|^2&=\frac{4s^2}{(N_C^2-1)}\left(\frac{g_s^2C_F}{|\mathbf{p}_1|^2}\right)\times\left(\prod_{i=2}^{n-1}\frac{4g_s^2C_A}{|\mathbf{p}_i|^2}\right)\times\left(\prod_{i=1}^{n-1}e^{2\alpha(\mathbf{q}_{i})\left(y_{j+1}-y_j\right)}\right)\times\left(\frac{g_s^2C_F}{|\mathbf{p}_n|^2}\right).
\end{split}
\end{align}
This expression makes full use of the simplifications allowed by the MRK region eq. \ref{eq:MRK}. The only real-emission corrections which are relevant to LL order are emissions of gluons within the rapidity span of the outgoing quarks. Virtual corrections exponentiate in this limit and are described by the Regge trajectory of the associated $t$-channel momentum $\mathbf{q}_i$,
%At leading order, this allows for the maximal number of $t-$channel exchanges of gluons. %A one-loop virtual correction to such a $t-$channel exchange is described by the factor
\begin{align}
	\label{eq:trajectory}
	\begin{split}
	\alpha(\mathbf{q}_{i})=-g_s^2C_A\frac{\Gamma(1-\epsilon)}{(4\pi)^{2+\epsilon}}\frac{1}{\epsilon}\left(\frac{\mathbf{q}_{i}^2}{\mu^2}\right)^\epsilon.
	\end{split}
\end{align}
The IR singularity here can be regularised by the procedure described in \cite{Andersen:2011hs}.
\par
%Processes with incoming gluons are described by a simple replacement of $C_F\to C_A$. 
While expressions such as eq. \ref{eq:LLMRK} are useful for probing the asymptotic limit of QCD, the kinematic approximations are too severe to be directly applied to the LHC. This motivates constructing a framework which maintains the logarithmic accuracy of eq. \ref{eq:LLMRK} while relaxing many of the kinematic approximations used to obtain it.
\section{The High Energy Jets framework}
\label{sec:HEJ}
%In the previous section we discussed the simple structure of QCD amplitudes at LL order. 
%The expressions quoted in the previous section were derived taking full advantage of the relevant %kinematic limits. Such simplifications are allowed within the quoted logarithmic accuracy, and are %useful if e.g. phase-space integrals are to be performed analytically.
%On the other-hand, 
%If we utilise Monte Carlo event generation to perform the phase-space integration, we are not forced to make such extreme kinematic approximations and so retain much more information about the fixed-order physics. 
%Utilising Monte Carlo event generation also enables 
%It also allows the use of arbitrary phase-space cuts allowing for direct comparison with experimental analyses. 
%\par
%In the following subsection we introduce the HEJ framework at LL accuracy. We then describe how this %framework may be extended to NLL accuracy.
\subsection{HEJ at LL}
We begin by noting that at LO, the amplitude for $qQ\to qQ$ scattering already has the form of high-energy factorisation without taking any kinematic approximations:
\begin{equation}
	\label{eq:M0qQtoqQ}
	i\mathcal{M}^{(0)}_{qQ\rightarrow qQ}=
	\left( i g_s T_{1a}^{C} j_{q \to q}^\mu(p_a, p_1)\right)
	\left( \frac{-i}{t} \right)
	\left( i g_s T_{2b}^{C}  j_{Q \to Q \ \mu}(p_b, p_2) \right),
\end{equation}
where we have defined the factorised spinor current
\begin{align}
	\label{eq:jqq}
	\begin{split}
	j_{q \to q}^\mu(p_1,p_a)&=\bar{u}^{\lambda_1}(p_1)\gamma^{\mu} u^{\lambda_a}(p_a).
	\end{split}
\end{align}
If we use these quark currents as our factorised building blocks we can construct a framework which is LO accurate for this process without loosing LL accuracy \cite{Andersen:2009nu}.
%upon the further approximation
%\begin{equation}
%    j_{q \to q}^\mu(p_1,p_a)j_{Q \to Q \ \mu}(p_b, p_2)\xrightarrow[\mathrm{MRK}]{}4s
%\end{equation}
%In Feynman gauge, 
The leading-order amplitude for $qg\to qg$ contains
%of diagrams with 
$s$- and $u$-channel diagrams as well as a $t$-channel diagram 
%so one might expect that a similar approach will not work for this process. 
However in ref.~\cite{Andersen:2009he} it was shown that for the dominant helicity configurations, the $qg\to qg$ amplitude can be written exactly in the same form as $qQ \to qQ$ with $C_F$ replaced with a momentum-dependent factor $K_g$.
%the replacement
%\begin{equation}
%\label{eq:CAM}
%C_F\to K_g(p_b^-,p_2^-)\equiv \frac{1}{2}\left(C_A- %\frac{1}{C_A}\right)\left( %\frac{p_b^-}{p_2^-}+\frac{p_2^-}{p_b^-}\right)+\frac{1}{C_A}.
%\end{equation}
%This colour-kinematic factor agrees with the simple colour-factor replacement mentioned above, with the approximation $p_b^-\approx p_2^- $, valid in the MRK limit.
%This colour-kinematic factor agrees with the well-known $C_F\to C_A$ replacement after the approximation $p_b^-\approx p_2^- $, valid in the MRK limit.
\par
The \HEJ form of LL amplitudes for pure-QCD processes is
\begin{align}
\label{eq:HEJLL}
\begin{split}
    \left|\mathcal{M}^{\mathrm{LL \ \HEJ}}_{f_af_b\to f_a\cdot (n-2)g\cdot f_b}\right|^2&=\frac{1}{4(N_C^2-1)}\left|j_{f_a\to f_a}\cdot j_{f_b\to f_b}\right|^2\left(g_s^2 K_{f_a}(p_a,p_1)\frac{1}{t_1}\right)\left(g_s^2 K_{f_b}(p_b,p_n)\frac{1}{t_{n-1}}\right)\\
    &\times\left(\prod_{i=1}^{n-1}e^{\omega(t_i)\left(y_{j+1}-y_j\right)}\right)\times\left(\prod_{i=2}^{n-1}\frac{-g_s^2C_A}{t_{i} t_{i+1}}\left|V_g(q_i,q_{i+q})\right|^2
    \right),
\end{split}
\end{align}
which should be compared to eq.~(\ref{eq:LLMRK}). While all-orders virtual corrections are described in the same manner, significantly more information about the emission of real radiation is contained via the \HEJ form of the so-called Lipatov vertex \cite{Lipatov:1976zz,Andersen:2009nu}
\begin{align}
\begin{split}
	V_{g}^{\rho}(q_i,q_{i+1})=-(q_i +q_{i+1})^{\rho}&+\frac{p_a^\rho}{2} \left(\frac{q_i^2}{p_{i+1}\cdot p_a} +\frac{p_{i+1}\cdot p_b}{p_a\cdot p_b}+\frac{p_{i+1}\cdot p_n}{p_a\cdot p_n} \right)+ p_a \leftrightarrow p_1\\
	& -\frac{p_b^\rho}{2} \left(\frac{q_{i+1}^2}{p_{i+1}\cdot p_b} +\frac{p_{i+1}\cdot p_a}{p_b\cdot p_a}+\frac{p_{i+1}\cdot p_1}{p_b\cdot p_1} \right)- p_b \leftrightarrow p_n.
\end{split}
\end{align}
%Alternative in terms of s_{ij}:
%\begin{align}
%\begin{split}
%	V_{g}^{\rho}(q_i,q_{i+1})
%	=-(q_i +q_{i+1})^{\rho}&+\frac{p_a^\rho}{2} \left(\frac{q_i^2}{s_{i+1,a}} %+\frac{s_{i+1,b}}{s_{a,b}}+\frac{s_{i+1,n}}{s_{a,n}} \right)+ p_a \leftrightarrow p_1\\&
%	 -\frac{p_b^\rho}{2} \left(\frac{q_{i+1}^2}{s_{i+1,b}} +\frac{s_{i+1,a}}{s_{ba}}+\frac{s_{i+1,1}}{s_{b1}} %\right)- p_b \leftrightarrow p_n.
%\end{split}
%\end{align}
%This is gauge invariant in all phase space, not just the MRK limit.
\par
The \HEJ framework can be extended to describe the production of a $W$, $Z$, or Higgs boson or same-sign $W$-pair production in association with jets~\cite{Andersen:2009nu,Andersen:2012gk,Andersen:2016vkp,Andersen:2018kjg,Andersen:2021vnf}. 
\subsection{NLL improvements to HEJ}
\label{sec:NLLbits}
The HEJ framework may be systematically improved by either matching to
fixed-order calculations (see section~\ref{sec:matching}), or by increasing the logarithmic accuracy of the
all-orders amplitudes. Leading powers
in $s/|t|$ in matrix elements lead to leading logarithms after integration.
Regge theory gives the scaling of matrix elements in terms of the spin of the
particles exchanged in the $t$-channels of the planar Feynman diagrams that can
be drawn for a given process when the legs are ordered in rapidity. A
flavour/momentum configuration contributes at LL if it permits an exchange of a
gluon in all $t$-channels. Beyond that, a flavour/momentum configuration
contributes at N$^m$LL if it permits an exchange of a gluon in all but $m$ of
the $t$-channels (see ref.~\cite{Andersen:2020yax} for further details).
\par
The first NLL component to be included in HEJ was the description of a gluon emission where that
gluon was more extreme in rapidity than one of the outgoing
quarks~\cite{Andersen:2017kfc}.  In ref.~\cite{Andersen:2020yax}, the necessary
pieces have been calculated to include all-order corrections to all configurations at 3$j$ and above whose leading contribution is at NLL order. This also includes the potential emission of a $W$ boson from these pieces. This is a gauge-invariant subset of the full NLL correction to
inclusive 2$j$ or $W+2j$ production. %These contribute to the improvement obtained in the HEJ2 NLO line in figure~\ref{fig:Final}.
\section{NLO matching}
\label{sec:matching}
We now describe the method used in ref. \cite{Andersen:2020yax} to increase the fixed-order accuracy of \HEJ to NLO.
%is will be useful to introduce some notation which compares the standard fixed-order approximation to a cross section with that of the all-orders approach of \HEJ.
We write the NLO calculation of the 2$j$ cross section as
\begin{equation}
\label{eq:NLO}
\sigma^{\text{NLO}}_{2j} =f_{2j}^{(2)}\alpha_s^2 + \left( f_{2j}^{(3)} +f_{3j}^{(3)} \right) \alpha_s^{(3)}.
\end{equation}
%This requires as input both the one-loop virtual corrections to the $n$-parton amplitude and the %leading order amplitude for ($n+1$)-parton production. 
Eq.~\eqref{eq:NLO} is finite, after cancellation of IR singularities between the virtual corrections and unresolved real emissions.
%\par
We may write the predictions of \HEJ as an all-orders expression in $\alpha_s$
\begin{equation}
\label{eq:HEJ}
\sigma^{\text{HEJ}}_{2j}=h_{2j}^{(2)}\alpha_s^2+(h_{2j}^{(3)}+h_{3j}^{(3)})\alpha_s^3+(h_{2j}^{(4)}+h_{3j}^{(4)}+h_{4j}^{(4)})\alpha_s^4+( h_{2j}^{(5)}+ h_{3j}^{(5)}+h_{4j}^{(5)}+h_{5j}^{(5)})\alpha_s^5+\mathcal{O}(\alpha_s^6).
\end{equation}
Within \HEJ \cite{Andersen:2019yzo}, $h^{(n)}_{nj}$ is matched to $f^{(n)}_{nj}$ where this is available. 
%We may truncate the predictions of \HEJ to any desired order in $\alpha_s$. For example,
We may truncate the above series to $\alpha_s^3$, which we will refer to as \HEJNLO:
\begin{align}
\begin{split}
\sigma^{\text{HEJ@NLO}}_{2j} &=h_{2j}^{(2)}\alpha_s^2 + (h_{2j}^{(3)} + h_{3j}^{(3)})\alpha_s^{3}. \\
\end{split}
\end{align}
In order to ensure that eq. \eqref{eq:HEJ} agrees with the fixed-order approach up to $\alpha_s^3$ without damaging the logarithmic accuracy of \HEJ, 
we multiply $\sigma^{\text{HEJ}}_{2j}$ by the ratio
\begin{align}
\frac{\sigma^{\text{NLO}}_{2j}}{\sigma^{\text{HEJ@NLO}}_{2j}} &=1+(f_{2j}^{(3)}-h_{2j}^{(3)}) \prod_{n=0}^{\infty}(-1)^n\alpha_s^{(n+1)}\frac{(f_{3j}^{(3)}+h_{2j}^{(3)})^n}{(f_{2j}^{(2)})^{(n+1)}}.
\end{align}
In the limit of $s\gg|t|$ this ratio tends to one, where in particular, $f_{2j}^{(3)} \to h_{2j}^{(3)}$.
%\par
Applying this reweighting factor to eq. \eqref{eq:HEJ} gives
\begin{align}
 \sigma^{\text{HEJ}}_{2j} \left( \frac{\sigma^{\text{NLO}}_{2j}}{\sigma^{\text{HEJ@NLO}}_{2j}} \right)&= f_{2j}^{(2)}\alpha_s^2+ \left(f_{2j}^{(3)}+f_{3j}^{(3)}\right)\alpha_s^3+ \left( h_{2j}^{(4)}+h_{3j}^{(4)}+f_{4j}^{(4)} \right) \alpha_s^4+\mathcal{O}(\alpha_s^5),
\end{align}
and we see that the resulting quantity is indeed NLO accurate in all of phase space.
\par
So far we have discussed cross-sections, but the same argument holds for any particular bin in a distribution of any observable in which we might be interested. For the practical implementation of this method, we must obtain separate histograms for NLO, \HEJ and \HEJNLO. In the study ref. \cite{Andersen:2020yax} we used Sherpa~\cite{Sherpa:2019gpd} with OpenLoops~\cite{Buccioni:2017yxi} to obtain pure NLO results. For \HEJNLO we truncate the all-orders prediction of the relevant \HEJ amplitudes. 
%This requires the truncation of the exponentiated virtual corrections (see e.g. eq. \ref{eq:HEJLL}) to first order in $\alpha_s$, and the limiting of real emissions to a single real gluon emission. 
%The organisation of the cancellation of the IR poles is performed with the same
%subtraction method as used in the all-orders calculation. 
We further separate the full \HEJ result into two pieces:
\begin{itemize}
    \item The LL resummed \HEJ predictions for 2$j$ inclusive (in practice we observe convergence after 6 jets) including LL resummation applied to the NLL configurations discussed in section~\ref{sec:NLLbits}, plus $3j$ configurations which are included at LO only via fixed-order matching.
    \item Configurations at $\ge4j$ which are included at LO only via fixed-order matching. 
\end{itemize}
This separation allows us to avoid applying NLO-matching to the fixed-order component of the HEJ result which begins at 4$j$. Finally, we can obtain the NLO-matched bin weight 
\begin{equation}
    w_{\mathrm{\HEJ} 2 \mathrm{ NLO}}=w_{\mathrm{\HEJ}( \ge 2j \ \mathrm{Resummed} + 3j \  \mathrm{FO})}\frac{w_{\mathrm{\HEJNLO}}}{w_{\mathrm{NLO}}}+w_{\mathrm{\HEJ}( \ge 4j \  \mathrm{FO})}.
\end{equation}
\begin{figure}[tb]
\centering
\begin{subfigure}{.5\textwidth}
  \centering
  \includegraphics[width=\linewidth]{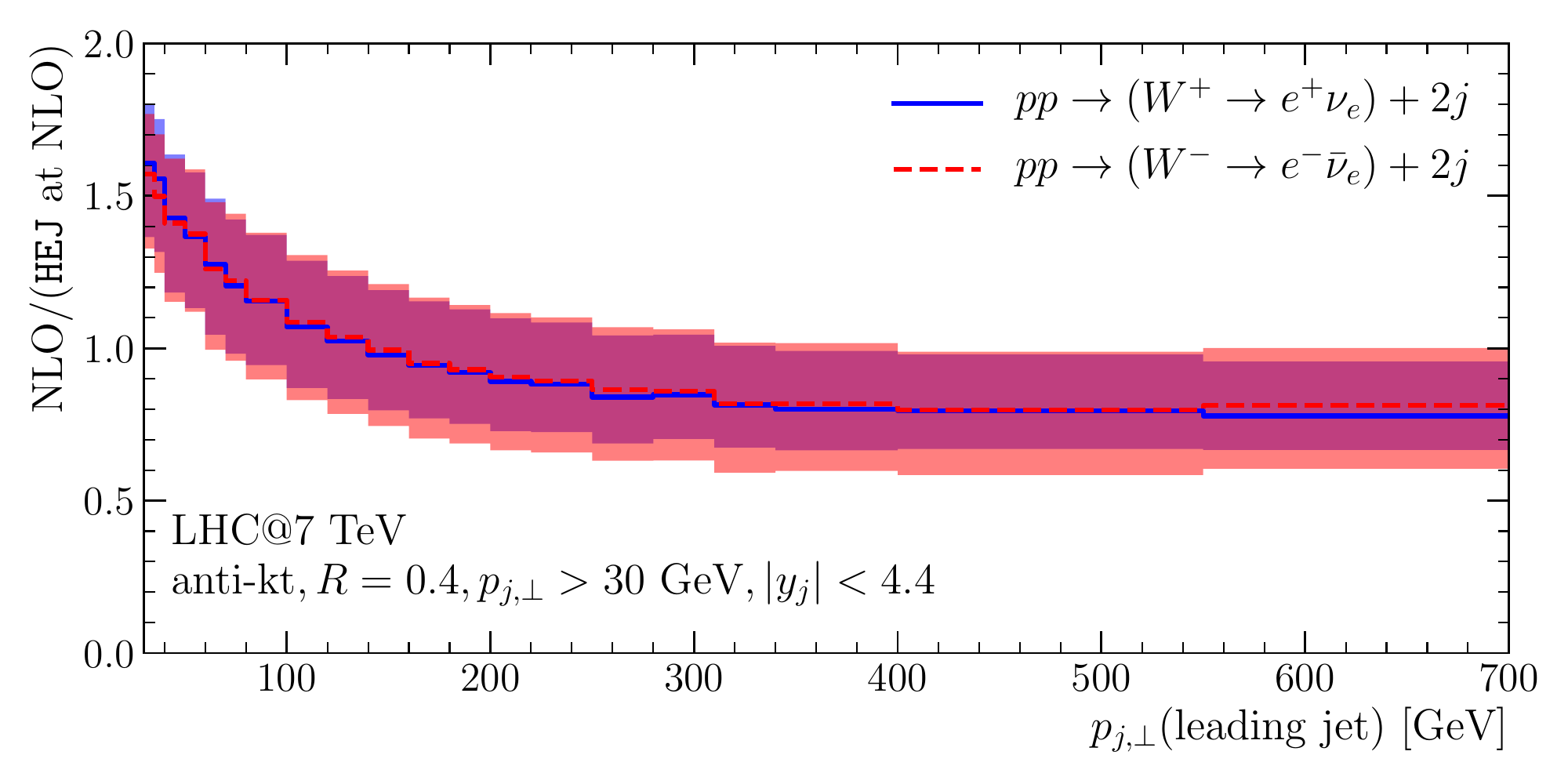}
%  \caption{Transverse
%  momentum of the
%      leading jet.}
  \label{fig:Reweighting_pt1}
\end{subfigure}%
\begin{subfigure}{.5\textwidth}
  \centering
  \includegraphics[width=\linewidth]{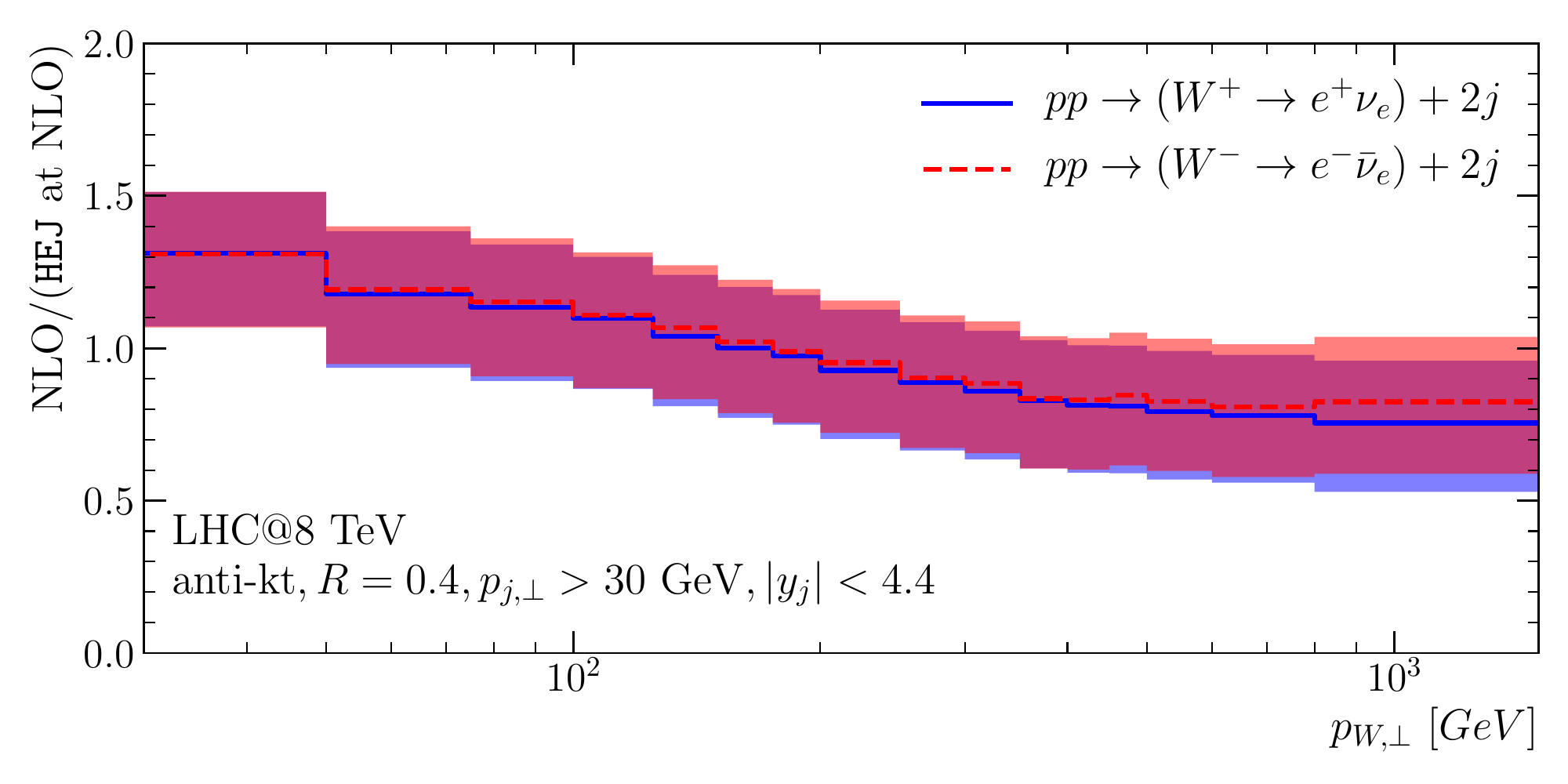}
%  \caption{Transverse
%  momentum of the
%      $W$ boson.}
  \label{fig:Reweighting_ptW}
\end{subfigure}
\caption{The NLO-matching corrections for distributions presented in \cite{Aad:2014qxa,ATLAS:2017irc} for inclusive $W^+ + 2j$ production (blue, solid) and $W^- + 2j$ production (red,
dashed).}
\label{fig:Reweighting}
\end{figure}
Fig.~\ref{fig:Reweighting} shows the matching corrections, ($w_{\mathrm{\HEJNLO}}/w_{\mathrm{NLO}}$), for two distributions corresponding to measurements in refs. \cite{ATLAS:2017irc,Aad:2014qxa}. Such $p_{\perp}$-based observables are of interest as they deviate from the transverse MRK condition (eq.~(\ref{eq:MRK})). As such, they are regions where we expect \HEJ to particularly benefit from matching to NLO. We apply the matching procedure separately for $W^+$ and $W^-$ production but observe no significant difference. All calculations presented use a central scale of $\mu_r=\mu_f=H_T/2$. The scale variation bands are obtained by varying $\mu_r$ and $\mu_f$ by a factor of two around this central scale, while keeping their ratio between 0.5 and 2. 
\par
\begin{figure}[bt]
\centering
\begin{subfigure}{.5\textwidth}
  \centering
  \includegraphics[width=\linewidth]{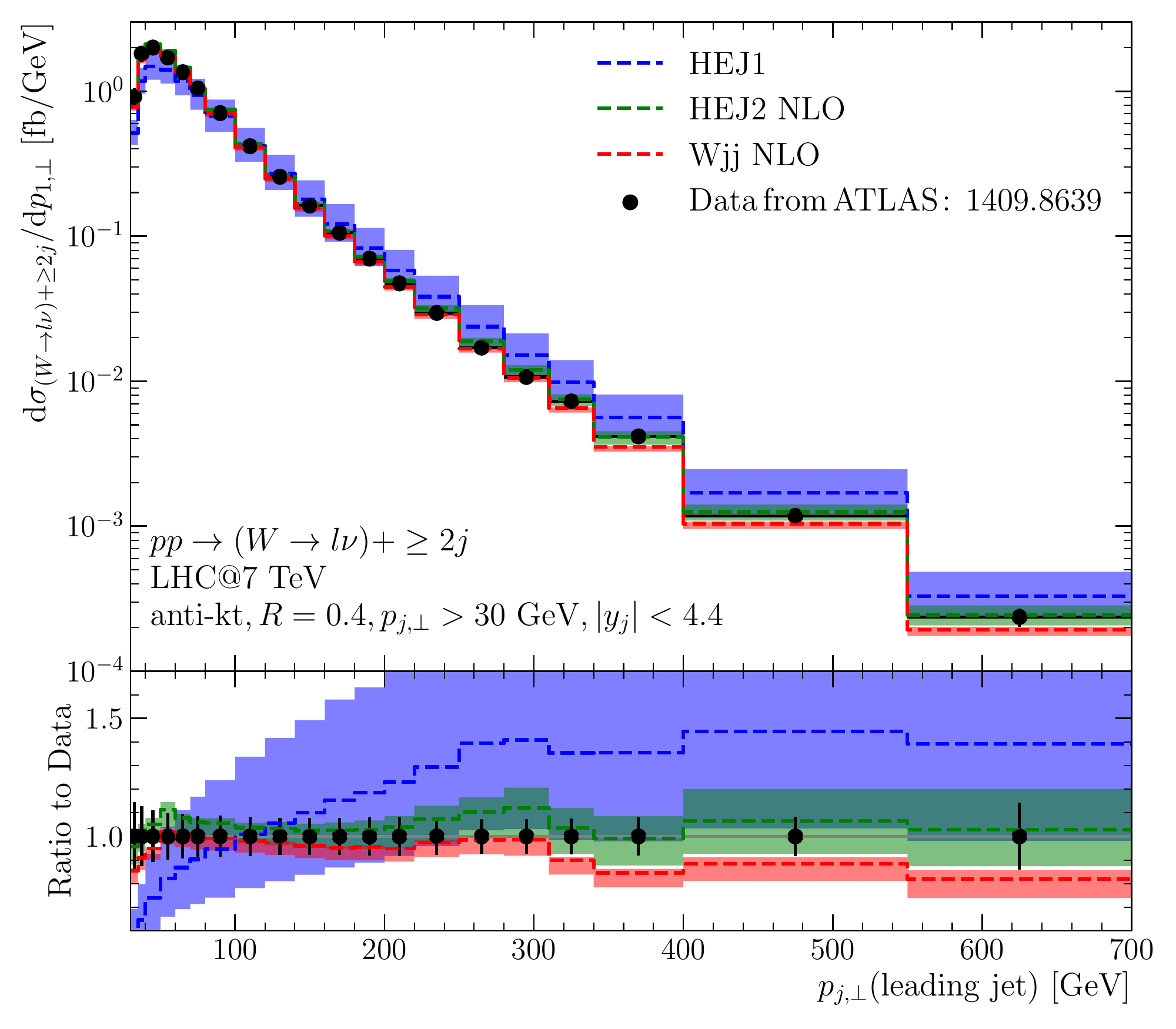}
%  \caption{Differential cross section for the transverse \\
%  momentum of the
%      leading jet.
%}
  \label{fig:Final_pt1}
\end{subfigure}%
\begin{subfigure}{.5\textwidth}
  \centering
  \includegraphics[width=\linewidth]{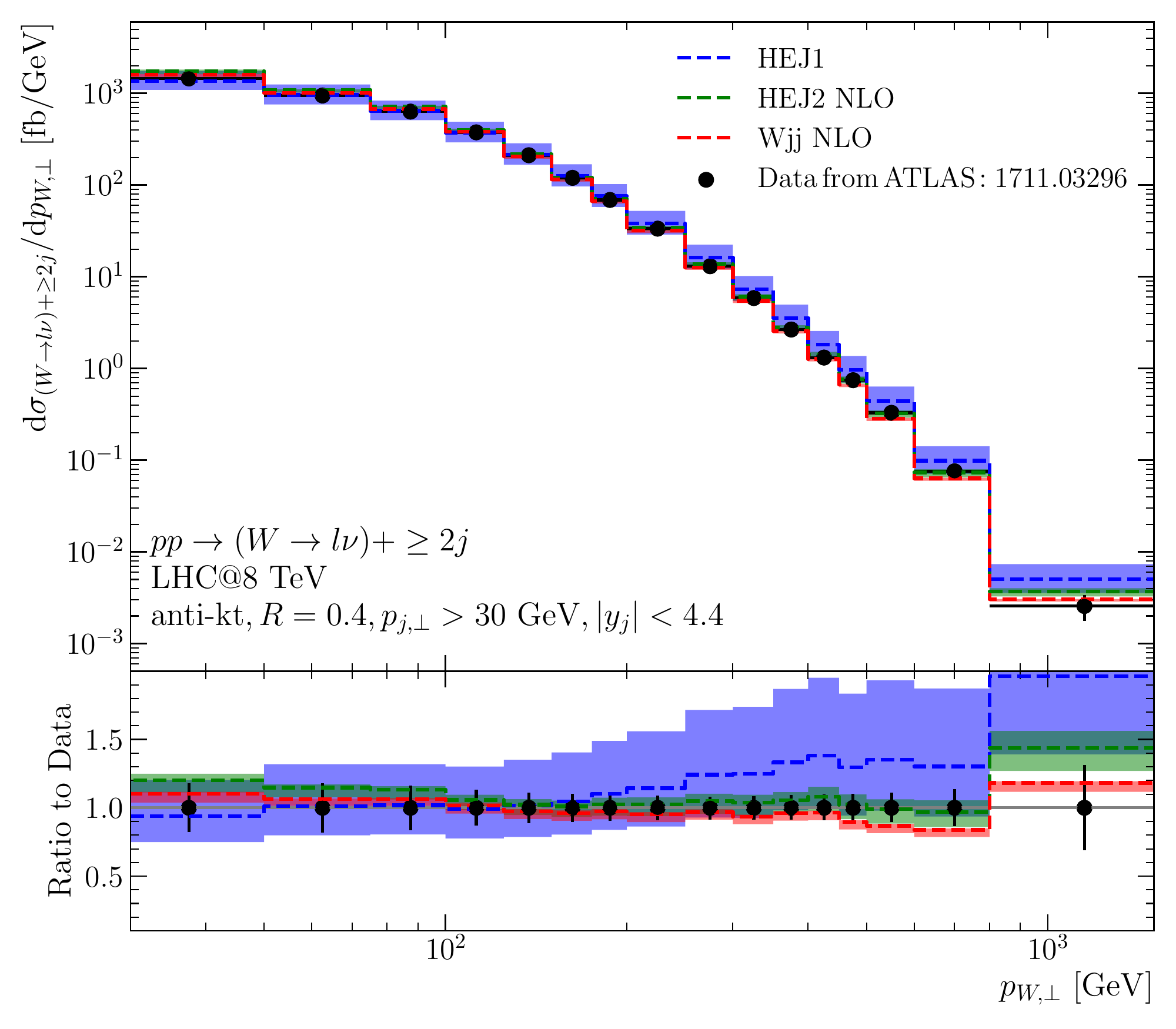}
%  \caption{Differential cross section for the transverse\\ 
%  momentum of the
%      $W$ boson.
%}
  \label{fig:Final_ptW}
\end{subfigure}
\caption{Predictions for $pp\to(W\to l\nu)+\ge2j$ for the LHC, compared to ATLAS data at 7 TeV~\cite{Aad:2014qxa} and 8 TeV~\cite{ATLAS:2017irc}. The content of the lines is discussed in the text. 
%The original \HEJone predictions are shown in blue, the %pure NLO predictions are shown in red and the new \HEJtwo %NLO predictions incorporating the methods of %sections~\ref{sec:NLLbits} and~\ref{sec:matching} are %shown in green.
}
\label{fig:Final}
\end{figure}

The application of these matching corrections are show in fig. \ref{fig:Final}. The blue line shows the results of the original LL \HEJ framework. The green line benefits from both the inclusion of the class of NLL improvements discussed in sec.~\ref{sec:NLLbits} and the application of the matching corrections shown in fig.~\ref{fig:Reweighting}. The pure NLO predictions are shown in red. We see that the NLO-matched \HEJ predictions give a favourable description of the data, in particular at large $p_{\perp}$ where NLO fixed-order and LL corrections are both significant. Furthermore, we see that the NLO-matched \HEJ predictions benefit from a significant reduction in the scale variation band.
\section{Conclusion}
In this proceeding we have introduced the High Energy Jets (\HEJ) framework, which provides all-orders predictions for jet production at the LHC that include high-energy effects and restore stability to the perturbative expansion.
As an example, we have applied this to the process $pp\to (W\to \ell \nu)+\ge2j$. We have seen that the combined effect of two improvements to the \HEJ framework, namely the inclusion of a gauge invariant class of NLL improvements and the matching to a NLO calculation, leads to close agreement with data. This highlights the importance of utilising the best of fixed-order and all-order calculations in order to provide precise and stable predictions for the LHC.
%(Add a brief summary on future directions of HEJ?)
%\clearpage
\section*{Acknowledgements}
We thank the organisers for an interesting and engaging workshop.  We would further like to thank the other members of the \HEJ collaboration for discussions during this work.  We gratefully acknowledge funding received from the ERC Starting Grant 715049 "QCDforfuture".

\bibliography{ISMD_HEJ.bib}

\nolinenumbers

\end{document}